\documentstyle[aps,preprint]{revtex}
\begin{document}
\draft
\tightenlines

\title{Compact directed percolation with movable partial reflectors} 

\author {Ronald Dickman$^{1,*}$ and Daniel ben-Avraham$^{2,\dag}$}
\address{
$^1$Departamento de F\'\i sica, ICEx,
Universidade Federal de Minas Gerais, Caixa Postal 702,
30161-970, Belo Horizonte - MG, Brasil\\
$^2$Physics Department and Center for Statistical Physics (CISP),
Clarkson University, Potsdam, New York 13699-5820}

\date{\today}

\maketitle
\begin{abstract}
We study a version of compact
directed percolation (CDP) in one dimension in which occupation  
of a site for the first time requires that a ``mine" or
antiparticle be eliminated.  This process is analogous to
the variant of directed percolation with a long-time memory, 
proposed by Grassberger, Chat\'e and Rousseau
[Phys. Rev. E {\bf 55}, 2488 (1997)] in order to understand spreading 
at a critical point involving an infinite number
of absorbing configurations.  The problem is equivalent to that of a pair 
of random walkers in the presence of movable partial reflectors. 
The walkers, which are unbiased, start one lattice spacing apart,
and annihilate on their first contact.
Each time one of the walkers tries to visit a new site, it is reflected 
(with probability $r$) back
to its previous position, while the reflector is simultaneously pushed 
one step away from the walker.  Iteration of the discrete-time
evolution equation for the probability distribution yields the
survival probability $S(t)$.  We find that $S(t) \sim t^{-\delta}$,
with $\delta$ varying continuously 
between 1/2 and 1.160 as the reflection probability varies between 0 and 1.  
\vspace{1em}

\noindent 
$^*${\small electronic address: dickman@cedro.fisica.ufmg.br } \\
$^{\dag}${\small electronic address: benavraham@clarkson.edu } \\


\end{abstract}


\section{Introduction}
 
Models that can become trapped in one of an infinite
number of absorbing configurations (INAC) exhibit 
unusual spreading dynamics at their critical point.
The most intensively studied model of this kind is the
pair contact process (PCP) \cite{pcp1,pcp2}.
INAC appears to be particularly relevant
to the transition to spatio-temporal chaos, as shown in a recent study
of a coupled-map lattice with `laminar' and `turbulent' states, which 
revealed continuously variable spreading exponents \cite{bohr}.

Anomalies in critical spreading for INAC (such as continuously variable
critical exponents) have been
traced to a long memory in the dynamics of the order parameter, $\rho$,
arising from a coupling to an auxiliary field that remains frozen in regions where
$\rho = 0$ \cite{inas,mgd}.  Grassberger, Chat\'e and Rousseau (GCR) \cite{gcr}
proposed that spreading in models with INAC could be understood
more easily by studying a model with a {\it unique} absorbing configuration,
but with a long memory of its initial preparation.

The GCR model is a variant of
bond directed percolation (DP) in which bonds connecting to
``virgin" sites (i.e., that have never been occupied), have a transmission
probability, $q$, that may differ from the value, $p$, for bonds 
to ``used" sites.  Used sites follow the usual DP rule.
If site $x$ has been occupied previously, then the 
probability that $x$ is occupied at time $t+1$ is $p$ if either $x-1$ or 
$x+1$ (but not both) are occupied at time $t$, $p(2-p)$ if both sites are 
occupied at time $t$, and
zero if neither is occupied.  
For virgin sites the parameter $p$ is replaced by $q$.
The dynamics begins (as in all spreading experiments) with activity restricted to a
small region of the lattice.  Grassberger et al. found in simulations that in
1+1 dimension, the critical point remains at $p_c = 0.644701$, the standard 
bond DP value \cite{essam88}, independent of $q$. 
They concluded that for $q < p_c$, the survival probability $S(t)$ 
decays faster than any power of $t$, at the critical point $p\!=\!p_c$.

In this work we study one-dimensional {\it compact} directed percolation 
(CDP) \cite{dk,cdp},
so called because gaps cannot arise within a string of occupied sites.  
Being exactly soluble, CDP provides a valuable test for ideas on scaling
in absorbing-state phase transitions.  For example, \'Odor and Menyhard 
recently found a continuously-variable survival exponent for CDP 
confined to a fixed parabolic region  \cite{odor00}.

The rules of
standard CDP are as for DP, described above, except that if both $x-1$ and $x+1$
are occupied at time $t$, then $x$ {\it must} be occupied at time $t+1$.  
(Note that CDP possesses two absorbing states: all vacant, and all occupied.)
If the process starts with only
a single occupied site, the state at any later time is specified by
the positions of a pair of random walkers, $w_1$ and $w_2$, which mark the 
extent of the occupied region.  
(Specifically, the occupied sites are: $w_1 \!+\! 1, w_1 \!+\! 2,..., w_2$.)
If we take the origin as the position of the 
original ``seed" particle, then $w_1 = 0$ and $w_2 = 1$ at $t=0$.  The 
stochastic evolution of $w_2$ is given by

\begin{equation}
w_2 (t+1) =  \left\{ \begin{array} {l}
 w_2 (t) + 1 \;\;\;\; \mbox{w.p. } p \\
 w_2 (t) - 1 \;\;\;\; \mbox{w.p. } 1\!-\!p
        \end{array} \right.
\end{equation}
while for $w_1$ the roles of $p$ and $1\!-\!p$ are interchanged.  Thus the 
length $Y(t) = w_2(t) - w_1(t)$ of the occupied region itself executes 
a random walk with transition probabilities: 

\begin{equation}
Y (t+1) =  \left\{ \begin{array} {l}
 Y (t) - 2 \;\;\;\; \mbox{w.p. } (1\!-\!p)^2 \\
 Y (t) \;\;\;\;\;\;\;\;\;\; \mbox{w.p. } 2p (1\!-\!p)   \\
 Y (t) + 2 \;\;\;\; \mbox{w.p. } p^2
        \end{array} \right.
\end{equation}
The state $Y=0$ (all sites vacant) is absorbing.  
In this work we focus on the case $p\!=\!1/2$ since for smaller
(larger) values $Y(t)$ is attracted to (driven away from) the
origin.  

Well known results on random walks \cite{barber} imply that for ($p\!=\!1/2)$,
the survival probability $S(t) \sim t^{-1/2}$, while the 
mean-square displacement, if the walker has not hit the origin up to 
time $t$, follows $\langle Y^2(t) \rangle_s \sim t$.  (The average is over
trials that survive until time $t$ or longer.)  The latter implies
an active region (in surviving trials) of extent $\sim t^{1/2}$,
so that the mean number of occupied sites, averaged over all trials,
is $n \sim t^0 $.  In the usual notation of absorbing-state phase
transitions \cite{torre,marro}, these results imply the exponent values
$\delta = 1/2$, $z=1$ and $\eta=0$.  (The exponents are defined via the
relations $P\sim t^{-\delta}$, $R^2 \sim t^z$, and $n \sim t^\eta$.)
These values satisfy the expected hyperscaling relation for a compact 
growth process in $d$ dimensions \cite{hypdk}:
\begin{equation}
\delta + \eta = \frac{dz}{2} \;.
\label{hypsc}
\end{equation}

Now we introduce a memory effect in CDP along the lines proposed
in Ref. \cite{gcr}.  Suppose that initially sites other than the
origin harbor static ``antiparticles" (or ``mines"), 
independently with probability $r$.  If site $x$ has a mine, then
the first particle to venture there is destroyed, and along with it 
the mine, so that in future, site $x$ can be occupied as in normal CDP.
In terms of the random walkers $w_1$ and $w_2$, a mine is effectively 
a reflecting boundary: the first time $w_1$ attempts to visit site 
$x$ (mined), it is reflected back to $x+1$, and at the same time
the reflector moves to $x-1$; similarly, $w_2$ will be reflected back to  
$x-1$ on its first visit to $x$, if it harbors a mine.
For $r>0$ our model represents the spread of activity into
a hostile environment, for example
the advance of a bacterial colony in a medium, with a 
preliminary contact facilitating expansion into new regions, or,
similarly, the spread of a political viewpoint in an initially
skeptical population.

We recently studied a simplified version of this
problem, involving a single random walker on the nonnegative
integers \cite{rwmpr}.  The walker is unbiased,
and starts at $x\!=\!1$, with $x\!=\!0$ absorbing.   The
reflector is initially at $x\!=\!2$.  On each visit to a new site, 
the walker is reflected with probability $r$,
and the reflector moves forward by one site.
Asymptotic analysis of the probability generating function
shows that the survival probability exponent varies continuously
with $r$: $\delta = (1\!+\!r)/2$.  In this work we analyze the two-walker
problem defined above, corresponding to a spreading CDP process.

The remainder of this paper is organized as follows.
In the following Section we show how CDP with reflectors can be
represented (despite the long memory) as a discrete-time 
Markov process.  We proceed to define an appropriate state space
and the associated transition probabilities.
In Sec. III we analyze the results of numerical iteration of
the probability evolution equations, yielding precision estimates
of the critical exponent $\delta$ and other asymptotic scaling
properties.  Sec. IV presents a summary and discussion.

\section{Model}

To investigate the scaling properties of CDP with reflectors, 
it is convenient
to enlarge the state space to include the positions of the reflectors;
this renders the process Markovian.  [The process $(w_1(t),w_2(t))$ is
evidently non-Markovian.]
We consider CDP in one dimension, starting from a
single active site.  The evolution of the active region is represented by the motion 
of a pair of unbiased random walkers, $w_1$ and $w_2$.  Each time a walker tries to jump to a 
new site it pushes a reflector ($R_1$ or $R_2$) to the left or right, and in the 
process the walker is reflected back with probability $r$; it remains at the
new position with probability $\overline{r} = 1\!-\!r$.  The generic
configuration is:
\[
\;\;\; \mbox{R}_1\;.\;.\;.\;.\;.\; \mbox{w}_1\;.\;.\;.\; \mbox{w}_2 \;
\;.\;.\;.\;.\;.\;.\;. \; \mbox{R}_2
\]
\noindent Due to translation invariance,
we require only three variables, $x$, $y$ and $z$, defined as follows:
\[
x = w_1 - R_1 - 1 ,
\]
\[
y = R_2 - w_2 - 1 ,
\]
\[
z = R_2 - R_1 - 2 .
\]
\noindent Then the distance between the walkers 
(i.e., the number of occupied sites),
is $z\!-\!x\!-\!y$; 
$x\!+\!y=z$ is the absorbing state.
We start with the walkers a unit distance apart, and the reflectors 
one lattice spacing away from the walkers, so that, initially,  
$z = 1$ and $x = y = 0$.
At each time step the walkers jump to the left or right with equal
probabilities.   $z$ is nondecreasing, with $1 \leq z \leq 2t + 1$, 
since the separation between 
reflectors can increase by at most two spacings at each step. 

Let $P(x,y,z;t)$ denote the probability of state $(x,y,z)$ at time $t$.
Transitions $(x,y,z) \to (x',y',z')$ may be grouped into three classes.
The simplest is for $x$ and $y$ both greater than zero.  Then $z$ cannot change,
since the walkers do not encounter the reflectors, and we have
$(x,y,z) \to (x',y',z)$ with $x'=x \pm 1$ and $y'= y\pm 1$; each of these
has a transition probability $W=1/4$.  Next consider $y>x=0$.  There are six
possible transitions, listed, along with their probabilitites, in Table I.
(The transition probabilities for $x>y$ are obtained by
noting that $W$ is symmetric under the simultaneous interchange 
of $x$ and $y$ and $x'$ and $y'$.)

\newpage
\begin{center}
\begin{tabular}{|c|c|c|l|}
\hline
$x'$ & $y'$ & $z'$ & $W$ \\
\hline
$0$ & $y\!+\!1$ & $z\!+\!1$ & $ \overline{r}/4 $ \\
$1$ & $y\!+\!1$ & $z\!+\!1$ & $ r/4 $ \\
$0$ & $y\!-\!1$  & $z\!+\!1$ & $ \overline{r}/4 $ \\
$1$ & $y\!-\!1$ & $z\!+\!1$ & $ r/4 $ \\
$1$ & $y\!+\!1$ & $z$        & $ 1/4 $ \\
$1$ & $y\!-\!1$ & $z$ & $ 1/4 $ \\
\hline
\end{tabular}
\label{tab1}
\end{center}
\centerline{Table I. Transition probabilitites for $y > x = 0$.}
\vspace{1em}

Finally, for $x=y=0$, there are eight possible transitions, as listed in Table II.

\begin{center}
\begin{tabular}{|c|c|c|l|}
\hline
$x'$ & $y'$ & $z'$ & $W$ \\
\hline
$0$ & $1$ & $z\!+\!1$ & $ \overline{r}/4 $ \\
$1$ & $0$ & $z\!+\!1$ & $ \overline{r}/4 $ \\
$1$ & $1$ & $z\!+\!1$ & $ r/2 $ \\
$0$ & $0$  & $z\!+\!2$ & $ \overline{r}^2/4 $ \\
$0$ & $1$  & $z\!+\!2$ & $ r\overline{r}/4 $ \\
$1$ & $0$ &  $z\!+\!2$ & $ r\overline{r}/4 $ \\
$1$ & $1$ &  $z\!+\!2$ & $ r^2/4 $ \\
$1$ & $1$ &    $z$      &  $1/4$        \\
\hline
\end{tabular}
\label{tab2}
\end{center}
\centerline{Table II. Transition probabilitites for $y = x = 0$.}
\vspace{1em}  

\noindent (Note that there are two distinct routes to the state $(1,1,z\!+\!1)$:
both walkers may jump to the left, with $w_1$ reflected back,
or both may jump to the right, with $w_2$ being reflected;
each of these events has a probability of $r/4$.)
Any move yielding $x'+y' \geq z'$ represents a transition 
into the absorbing state.
Starting from $P(x,y,z;0) = \delta_{z,1}\delta_{x,0}\delta_{y,0}$, 
we can iterate the above transition probabilities to find $P(x,y,z;t)$.  

In the three-variable representation,
the evolution is confined to an infinite wedge bounded by the planes $x\!=\!0$,
$y\!=\!0$, and $x\!+\!y \!=\! z$.  The latter plane is absorbing, while the first
two allow upward transitions (from $z$ to $z\!+\!1$ or $z\!+\!2$).  
Between vertical transitions, the process is confined to the triangle
$x \geq 0$, $y \geq 0$, $x\!+\!y \leq z$; away from the boundaries,
the evolution is that of an unbiased lattice walk with
steps between second-neighbor sites, on $Z^2$.

Suppose the process has just entered a given plane of constant $z$
from below.  Its continued survival is equivalent to the event that it
touches either the $x$ or the $y$ axis, and makes a further vertical
transition, {\it before} touching the line $x\!+\!y = z$.  Thus survival
of the process is related to the {\it splitting probabilities} for exiting a
two-dimensional triangular region via the different edges.  (Note that
the $x$ and $y$ axes are partly reflecting.)

The above transition probabilities define what we shall refer to as the
``two-step" model, in which both walkers jump at each time step, in
correspondence with the original CDP problem.  One may define a 
simpler ``one-step" model, in which only one of the walkers (chosen at
random, with equal likelihood) jumps at each step.  Since the set of
transitions is somewhat reduced (there are no transitions from 
$z$ to $z\!+\!2$, for example), this version would
appear to be more amenable to analysis.  We expect the two versions
to have identical asymptotic scaling properties.

The transition probabilities for the one-step process are as follows.
If $x$ and $y$ are both greater than zero, there are four possible
transitions,
$(x,y,z) \to (x\pm 1,y,z)$, and $(x,y,z) \to (x,y \pm 1,z)$,
each with probability $W=1/4$.  
For $y>x=0$, the five possible transitions are listed in Table III.

\begin{center}
\begin{tabular}{|c|c|c|l|}
\hline
$x'$ & $y'$ & $z'$ & $W$ \\
\hline
$0$ & $y$ & $z\!+\!1$ & $ \overline{r}/4 $ \\
$1$ & $y$ & $z\!+\!1$ & $ r/4 $ \\
$1$ & $y$ & $z$ & $ 1/4 $ \\
$0$ & $y\!+\!1$ & $z$ & $ 1/4 $ \\
$0$ & $y\!-\!1$ & $z$ & $ 1/4 $ \\
\hline
\end{tabular}
\label{tab3}
\end{center}
\centerline{Table III. One-step model: transition probabilitites for $y > x = 0$.}
\vspace{1em}  

Finally the transitions for the case $x \!=\! y \!=\! 0$ for the
one-step model are given in Table IV.

\begin{center}
\begin{tabular}{|c|c|c|l|}
\hline
$x'$ & $y'$ & $z'$ & $W$ \\
\hline
$0$ & $0$ & $z\!+\!1$ & $ \overline{r}/2 $ \\
$1$ & $0$ & $z\!+\!1$ & $ r/4 $ \\
$1$ & $0$ & $  z  $   & $ 1/4 $ \\
$0$ & $1$ & $z\!+\!1$ & $ r/4 $ \\
$0$ & $1$ & $  z  $   & $ 1/4 $ \\
\hline
\end{tabular}
\label{tab4}
\end{center}
\centerline{Table IV. One-step model: transition probabilitites 
for $x \!=\! y \!=\! 0$.}
\vspace{1em}

The evolution of the one-step process is again confined to the
wedge described above, and (in each plane) to the same triangular
region as the two-step process.  The principal differences are
that, away from the boundaries, the process corresponds to a
simple random walk with jumps between {\it nearest} neighbors,
and that all vertical transitions are from $z$ to $z\!+\!1$.

The probability distribution evolves via
\[
P(x,y,z;t\!+\!1) = 
\sum_{x',y',z'} W(x,y,z|x',y',z') P(x',y',z';t).
\]
Results from iteration of this equation are
discussed in Sec. III.

\subsection{Calculational scheme}

Given the symmetry of the transition probabilities
under exchange of $x$ and $y$ (and, simultaneously, of
$x'$ and $y'$), it follows that if we start from
a symmetric distribution, $P(x,y,z) = P(y,x,z)$, as is the case here,
then this property will be maintained throughout the
evolution.   This allows us to reduce the number of states by roughly
half: we need only study $x \leq y$.  The presence of the absorbing state
implies that $y \leq z\!-\!1$, and, therefore,
$0 \leq x \leq \min [y, z\!-\!y\!-\!1]$.

Since states with $x > y$ are not considered explicitly, we must
modify the iteration of the evolution equation as follows:
\vspace{1em}

\noindent (1) The contribution to $P(y',y',z';t\!+\!1)$ due to a transition 
$(x,y,z) \to (y',y',z)$ with $x \!<\! y$ should be {\it doubled}, to take into account
the corresponding contribution due to $(y,x,z) \to (y',y',z)$, which is not
represented explicitly in the dynamics. Similarly, a transition from 
$(x,y,z)$ to the absorbing state should have its weight doubled, if $x<y$.
\vspace{1em}

\noindent (2) In a transition $(x,y,z) \to (x',y',z')$, with $x < y$ {\it and} $x' > y'$, the
contribution should instead be added to $P(y',x',z';t\!+\!1)$, to include the 
mirror process, which, again, is not represented explicitly.
\vspace{1em}

\noindent These rules are summarized in Table V, which gives the weights
associated with each transition, given that states with $x >y$ are not represented
explicitly.

\begin{center}
\begin{tabular}{|c|c|}
\hline
From $x\!=\!y$ to & weight \\
\hline
$x'=y'$ & 1 \\
$x'<y'$ & $ 1 $ \\
$x'>y'$ & $ 0 $ \\
absorbing & $ 1 $ \\
\hline
From $x\!<\!y$ to & weight \\
\hline
$x'=y'$ & 2 \\
$x'<y'$ & $ 1 $ \\
$x'>y'$ & 1 for $(y',x')$ \\
absorbing & $ 2 $ \\
\hline
\end{tabular}
\end{center}
\label{tab5}
\centerline{Table V. Transition weights.}
\vspace{1em}

\noindent The zero entry for $x\!=y\! \to x'\!>\!y'$ 
means that such transitions are ignored.

\section{Results}

We have iterated the discrete-time
evolution equation derived above numerically.
To iterate the two-step process for $t_m = 2000-5000$ time steps, one
requires values of $z$ of up to 200 - 320, and of $x$ and $y$
up to 110 - 230, depending on $r$.  (The larger $r$ is, the less
rapidly the process spreads, and the smaller the arrays need be.
The required size scales, naturally, as $\sqrt{t_m}$.  For $t_m=2000$
the iteration requires about 20 min. to 1 hour of cpu time on an alpha
workstation.)   For the one-step process we use an upper limit of
250 for all three variables, which proves more than sufficient for
$t_m = 2000$.
The CDP plus reflectors problem is, of course,
easily studied via Monte Carlo simulation.  But we have found that 
numerical iteration furnishes an order of magnitude higher
precision than direct simulation, for the same expenditure of cpu time.

For each $r$ we calculate the survival probability $S(t)$, the
first and second moments, $\langle Y \rangle_t$ and 
$\langle Y^2 \rangle_t$ of the extent of the active region 
(i.e., the distance between the walkers, $z\!-\!x\!-\!y$), and the
probability distribution $P(Y)$ at $t_m$.  
(The distribution and moments of $Y$ are taken over the surviving
sample at time $t$.  With the array sizes mentioned above we determine
the quantities of interest to a precision of better than one part in 
$10^6$.)

$S(t)$ and the moments of $Y$ are found to follow power laws,

\begin{equation}
S(t) \sim t^{-\delta}
\end{equation}
\begin{equation}
\langle Y \rangle_t \sim t^{\eta_s}
\end{equation}
\begin{equation}
\langle Y^2 \rangle_t \sim t^{2\eta_s}
\end{equation}
(Here the subscript `s' denotes surviving sample; the
exponent $\eta = \eta_s - \delta$.)

Precise estimates of the exponents are obtained by studying 
local slopes, for example $\delta(t) \equiv d \ln P/d \ln t$, and
similarly for the other exponents.  Based on experience with the 
random walk with movable partial reflectors \cite{rwmpr}, we expect a generic 
correction to scaling exponent of 1/2; we therefore plot the local slopes 
versus $t^{-1/2}$.  Such plots (see Fig. 1), are roughly linear, but show 
a certain degree of curvature, indicating (as is to be expected) that 
corrections of order $t^{-1}$ are still significant.  The local slope data 
are fit nearly perfectly by a quadratic form in $t^{-1/2}$;  the intercept
yields our estimate for the critical exponent.  
An exception is the case $r\!=\!1$, for which the correction to scaling
exponent appears to be 1.  (This may be seen explicitly in the
case of a single random walk with a partial movable reflector \cite{rwmpr}.)
For $r\!=\!1$ we derive our estimate for $\delta$ from an analysis of
the local slope as a function of $t^{-1}$.  The extrapolated values for
$\delta$ are very stable under changes in the interval used
(e.g., $t^{-1/2} < 0.1$, or $t^{-1/2} < 0.04$), and in $t_m$ (2000 or
5000 time steps).  We estimate the uncertainty of extrapolation
as $\leq 2 \times 10^{-4}$.  This is supported by our results for
the one-step process: the estimates for $\delta$ differ from those
for the two-step model by at most 0.0005.  (For $r=1$ for example,
we find $\delta = 0.1597$, compared with $\delta = 0.1595$
in the two-step case.)

The analysis described above yields $\eta_s = 1/2$ to within one part 
in 5000.  Thus the only independent exponent is $\delta$. 
Our results for $\delta(r)$ are given 
in  Table IV. 
As shown in Fig. 2, $\delta$ appears to vary linearly with $r$.  
A simple linear expression, 
\begin{equation}
\delta = \frac{1}{2} + \frac{2r}{3}, 
\label{conj}
\end{equation}
reproduces the
data to within 6 parts in 1000.  The simplicity of this expression,
and its similarity to the single-walker result, $\delta = 1/2 + r/2$,
lead us to adopt Eq. (\ref{conj}) as a conjectured exact formula.
Analysis using least-squares fitting suggests, however, that $\delta(r)$
is weakly nonlinear.  We obtain an excellent fit to our data using an 
expression of the form,
\begin{equation}
\delta = \frac{1}{2} + a r + b r^2 ,
\label{dells}
\end{equation}
with $a = 0.6434$ and $b=0.0167$.  The typical errors associated with a purely
linear fit are on the order of $2 \times 10^{-3}$, about an order of magnitude
larger than the uncertainty in $\delta$.  
(The typical error for the best quadratic fit is about $4 \times 10^{-4}$.)
The nonlinear dependence of $\delta$ on $r$
therefore appears to be real, not just an effect of finite numerical 
precision.   
It is conceivable, nevertheless, that corrections to scaling
introduce systematic errors in the numerical analysis, giving rise
to apparent nonlinearities.  We defer the verification of Eq. (\ref{conj}),
which would appear to require either an analytic solution or improved
numerics, to future work.

\begin{center}
\begin{tabular}{|c|c|c|c|}
\hline
$r$ & $\delta$ &   $A$  & $m$  \\
\hline
$0$ & $1/2$    & 1.7724 & 1.2732  \\
0.1 & 0.5642   & 1.6508 & 1.2679  \\
0.2 & 0.6289   & 1.5494 & 1.2628  \\
0.3 & 0.6940   & 1.4633 & 1.2580  \\
0.4 & 0.7598   & 1.3892 & 1.2539  \\
0.5 & 0.8259   & 1.3246 & 1.2502  \\
0.6 & 0.8924   & 1.2678 & 1.2469  \\
0.7 & 0.9591   & 1.2174 & 1.2439  \\
0.8 & 1.0258   & 1.1722 & 1.2411  \\
0.9 & 1.0927   & 1.1316 & 1.2386  \\
1.0 & 1.1595   & 1.0947 & 1.2362  \\
\hline
\end{tabular}
\end{center}
\label{tab6}
\centerline{Table VI. Numerical results.}
\vspace{1em}

Table VI also contains results for the amplitude $A$ of
the mean activity, defined via
\begin{equation}
\langle Y \rangle \simeq A t^{1/2},
\label{defA}
\end{equation}
and for the asymptotic 
moment ratio 
\begin{equation}
m = \lim_{t \to \infty} \frac{\langle Y^2 \rangle_t}
{   \langle Y \rangle_t^2} \;.
\label{defm}
\end{equation}
The amplitude decreases smoothly with $r$ as shown in Fig. 3.
$m$ is a measure of the shape of the position distribution.
For $r=0$ our numerical estimate is consistent with $4/\pi = 1.27324...$,
as expected for Brownian motion on the line with the origin reflecting,
for which the asymptotic probability density is 
$P_Y(x) = (x/\sigma^2) \exp[-x^2/2\sigma^2]$, with $\sigma^2 = t$.
The ratio decreases steadily with $r$, but not by very much (see Fig. 3), 
showing that the random walk result still serves as a reasonable 
approximation for $r>0$.  The moment ratio $m$ takes the same values
in both versions of the process, confirming that it is a
universal quantity.

The effect of the reflectors is clearly
evident in the distribution in the number of active sites,
$P(Y)$.  Fig. 4 compares
$P(Y)$ for $r\!=\!0$, 0.5, and 1, (for $t= 2000$ time steps, two-step
process),
showing that the distribution shifts to smaller $Y$ values with
increasing reflection probability $r$.  (Note that for $r\!=\!0$,
$P(Y,t)\!=\!0$ for $Y\!+\!t$ odd.  We have therefore multiplied the
distribution for $r\!=\!0$ by one half, to facilitate comparison
with the other cases.)  Despite the changes in form, the tail of the distribution
remains Gaussian in all cases.

\section{Discussion}

We have studied spreading in compact directed percolation on a 
one-dimensional lattice at its critical point, modified so that
activation of virgin sites is less probable than reactivation of
a previously active site.  The problem is equivalent
to a pair of random walkers subject to movable partial 
reflectors.   Two variants are considered: the two-step process,
in which both walkers move at each step, and a one-step process,
in which, at each step, only one walker (selected at random) jumps.
We study these processes via exact numerical iteration
of the probability distribution for finite times ($\leq 5000$ time steps).

We find that the survival probability critical exponent 
$\delta$ varies continuously with $r$. Our results
indicate a weak nonlinearity in the function $\delta(r)$, despite
the fact that the data are rather well represented by a simple
linear expression, Eq. (\ref{conj}).
Since, in the case of a single random walker subject to a partial movable 
reflector \cite{rwmpr}, we found a strictly linear dependence of the survival probability 
exponent on $r$, this nonlinearity is somewhat surprising.  On 
the other hand, the present problem is related to splitting 
probabilities on a two-dimensional domain (rather than on the
line, as is the case for a single walker), allowing for a more complicated
functional dependence. 

Our finding of a continuously-variable survival probability exponent is 
fully consistent with previous results for the single walker.  One 
may, moreover, understand the fact that $\delta$ increases more rapidly 
with $r$ than for the single walker, since in CDP the spreading process 
feels the effects of {\it two} reflectors.  In the absence of an exact 
analysis or rigorous argument, however, we have no qualitative 
understanding of the values for $\delta(r)$ that we have found numerically.  
This question, and the verification of Eq. (\ref{conj}), 
remain as interesting challenges for future work.
\vspace{1em}

\noindent {\bf Acknowledgements}
\vspace{1em}

We are grateful to Deepak Dhar for stimulating discussions.
This work was supported by CNPq, Brazil and the NSF, USA.


\newpage
\noindent {\bf Figure Captions}
\vspace{1em}

\noindent FIG. 1. Inset: survival probability $S(t)$ for the two-step
process, $r\!=\!0.8$.  Main graph: local slope $\delta(t)$ versus
$t^{-1/2}$ for the same system.  `$\times$' on $y$-axis: (upper)
extrapolated value ($\delta = 1.0256$); (lower) $\delta = 1.0333$ 
predicted by Eq. (\ref{conj}).
\vspace{1em}

\noindent FIG. 2. Survival exponent $\delta(r)$ from
iteration of probability distribution (points).  The solid line is
a quadratic least-squares best-fit to the data.  Inset:
best-estimate for $\delta$ less the value predicted by Eq. (\ref{conj}).
\vspace{1em}

\noindent FIG. 3. Upper panel: amplitude $A$ of the mean activity
as a function of $r$;
lower: moment ratio $m$ of the activity distribution.
\vspace{1em}

\noindent FIG. 4. Probability distribution $P(Y)$ of the activity
in the two-step model (conditioned on survival) after 2000 time steps,
for (left to right) $r\!=\!1$, 0.5, and 0.
\vspace{1em}

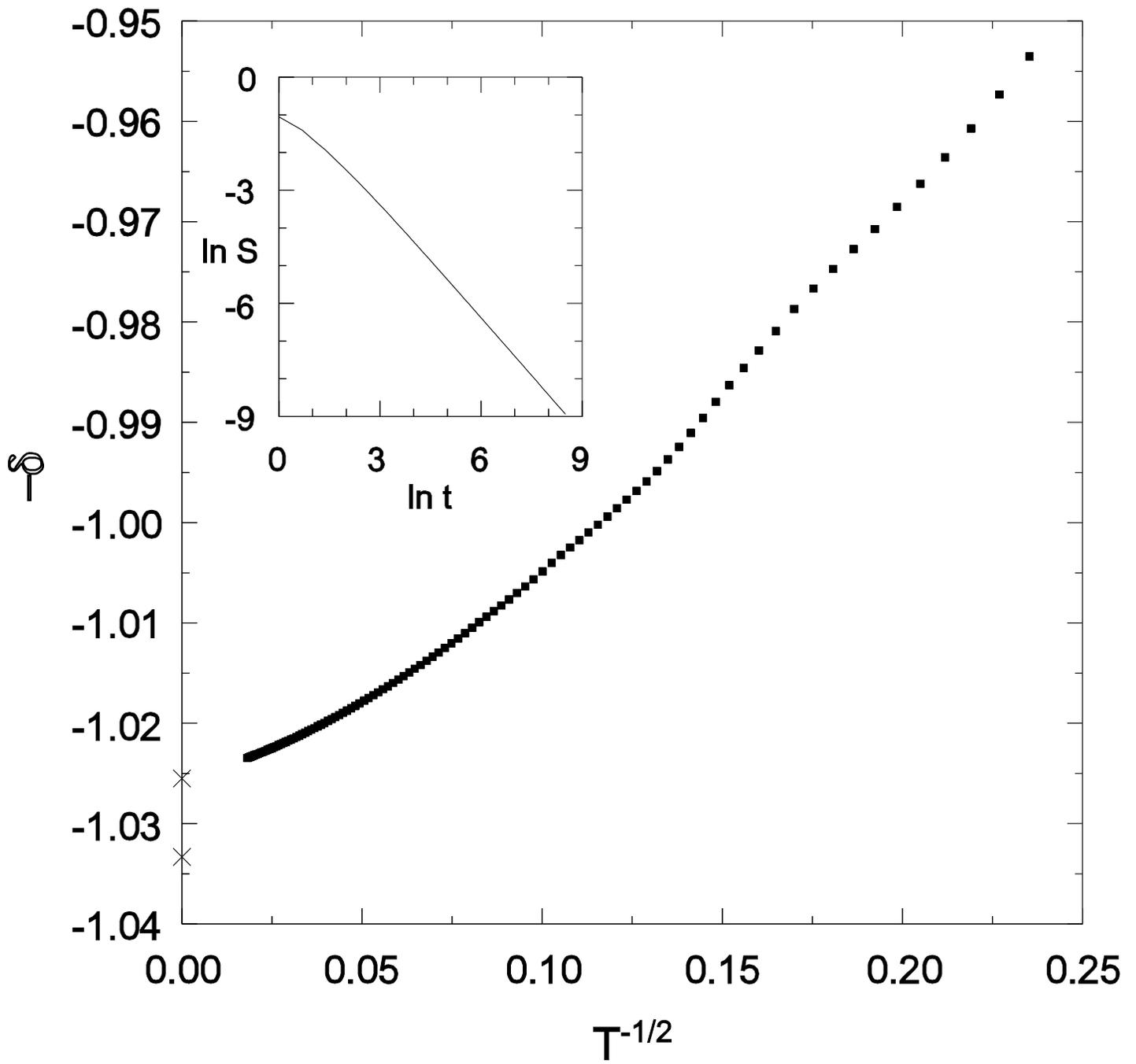
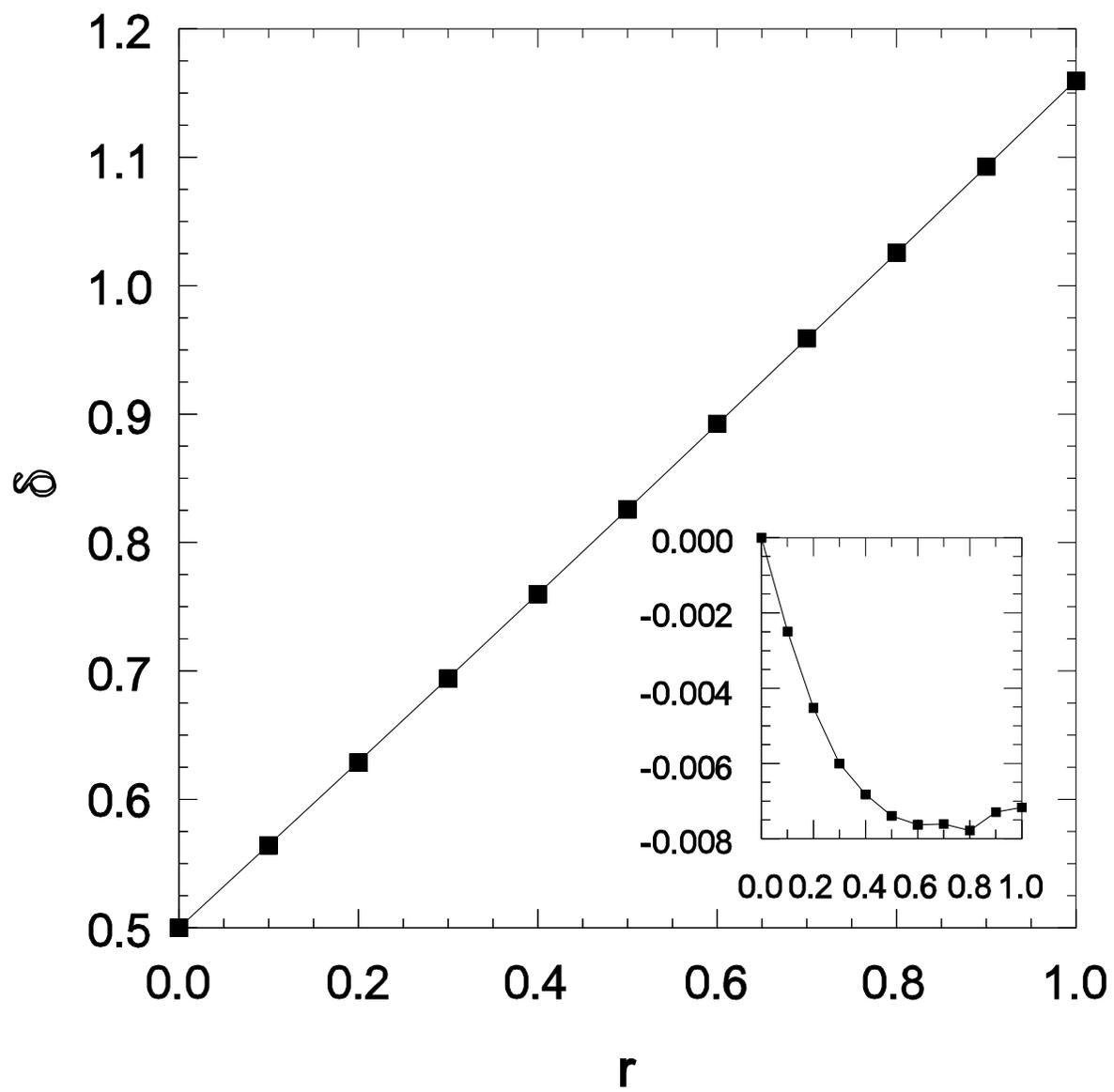
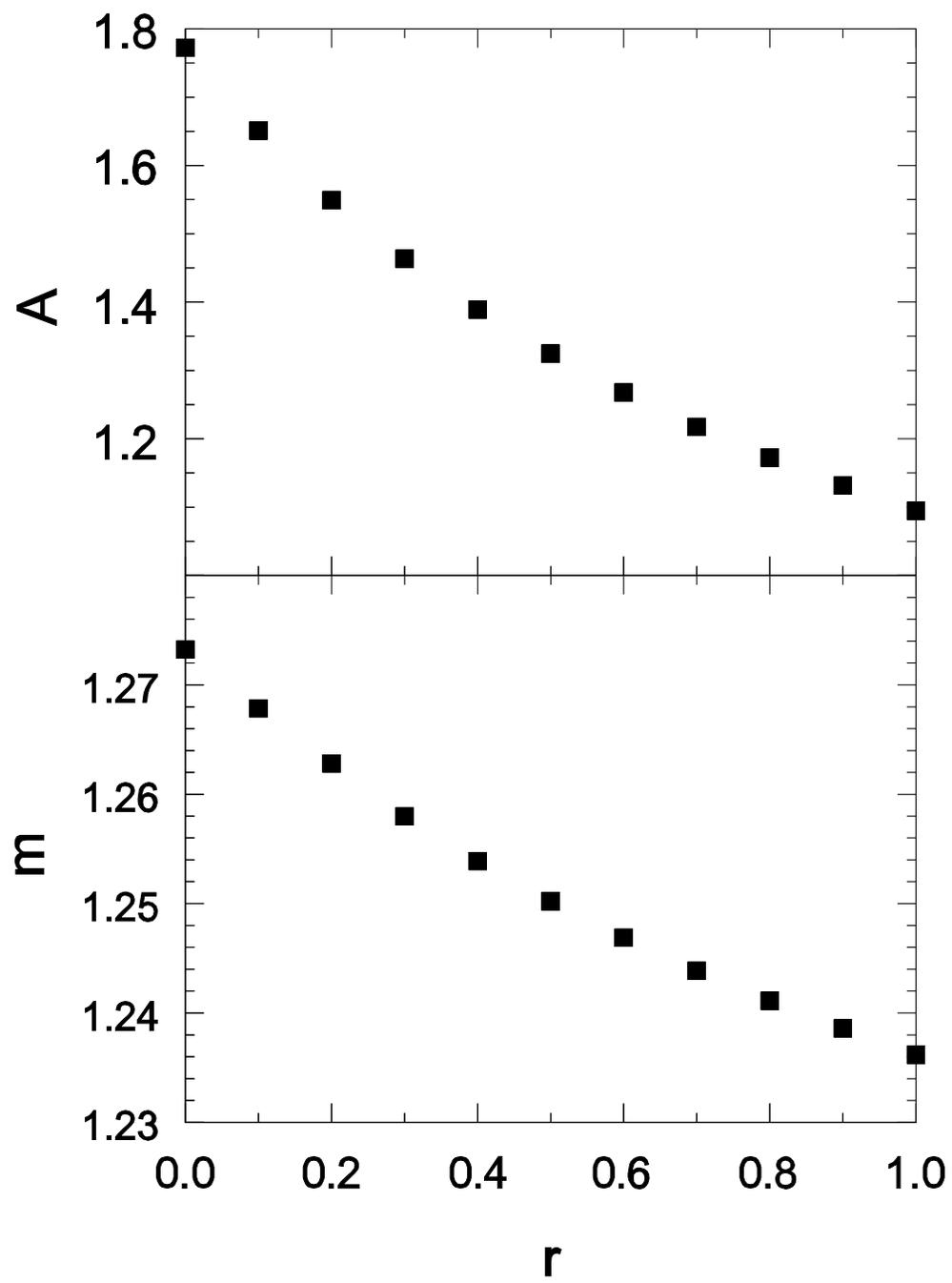
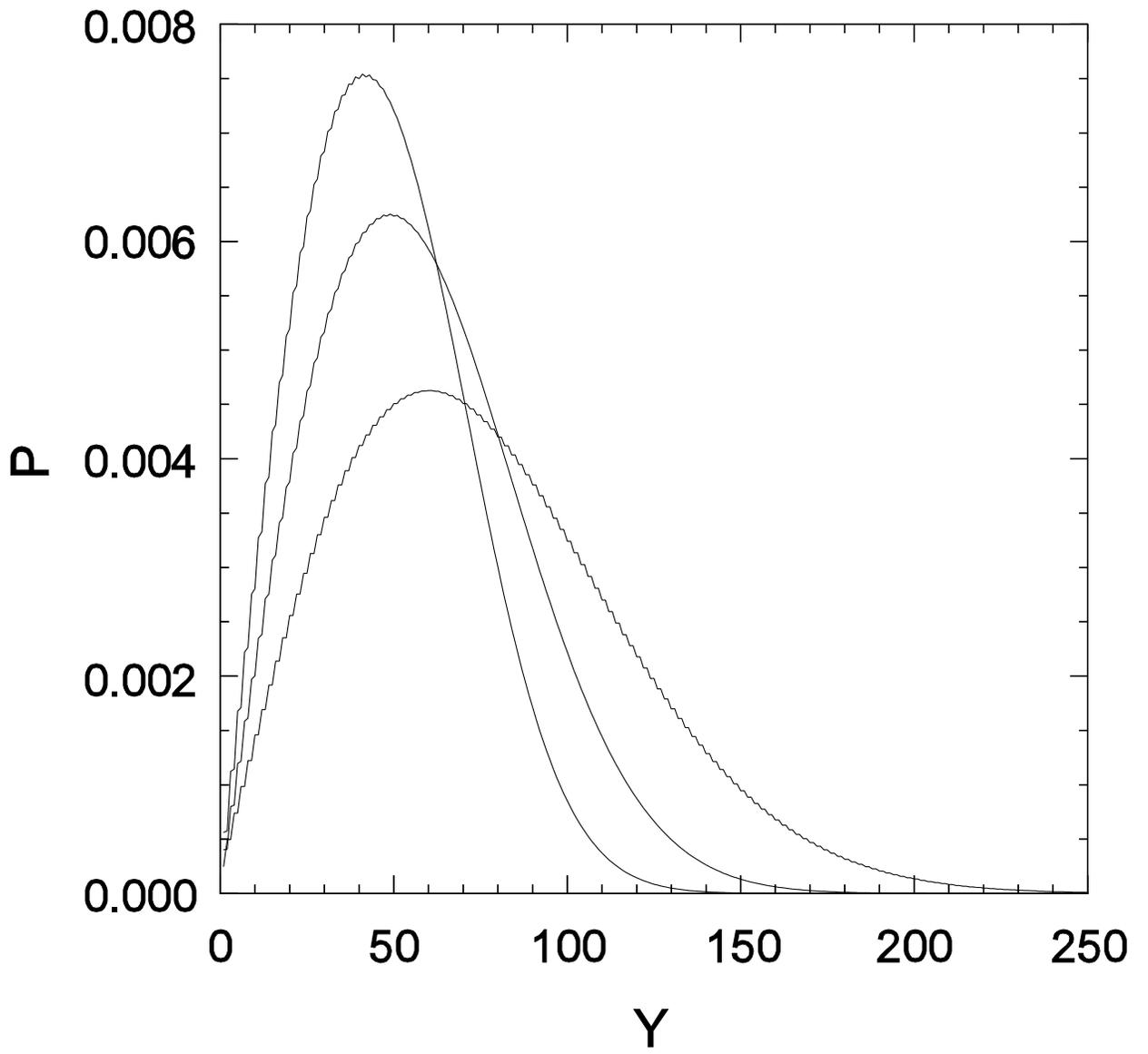
\end{document}